\documentclass[10pt,conference]{IEEEtran}
\IEEEoverridecommandlockouts

\usepackage{cite}
\usepackage{amsmath,amssymb}
\usepackage{graphicx}
\usepackage{booktabs}
\usepackage[hidelinks]{hyperref}
\usepackage{balance}
\usepackage{orcidlink}

\begin{document}

\title{Context Is a Pipeline, Not a Window:\\
Information Attrition in LLM Agent Harnesses\thanks{Replication package:
\url{https://doi.org/10.5281/zenodo.23614755}}}
\author{
\IEEEauthorblockN{Yuvraj Sehgal\,\orcidlink{0009-0002-1980-604X}}
\IEEEauthorblockA{\textit{Department of Computer Science}\\
\textit{Brock University}\\
St.\ Catharines, Ontario, Canada\\
ys19rk@brocku.ca}
\and
\IEEEauthorblockN{Mahsa Panahandeh\,\orcidlink{0000-0002-6369-8982}}
\IEEEauthorblockA{\textit{School of Electrical Engineering}\\
\textit{and Computer Science}\\
\textit{University of Ottawa}\\
Ottawa, Ontario, Canada\\
mpanahan@uottawa.ca}
\and
\IEEEauthorblockN{Naser Ezzati-Jivan\,\orcidlink{0000-0003-1435-6297}}
\IEEEauthorblockA{\textit{Department of Computer Science}\\
\textit{Brock University}\\
St.\ Catharines, Ontario, Canada\\
nezzatijivan@brocku.ca}
}

\maketitle

\begin{abstract}
LLM agents are judged by two numbers: the size of the context window, and
how often the final answer is correct. Both numbers miss a third thing: how
much of a tool's output is lost before the model ever sees it. We measured
this loss in a real tool-using agent. The agent finds the cause of injected
faults by reading Linux kernel traces. The context window was almost empty:
across $720$ runs, the single largest prompt used only $5.75\%$ of it. Yet a
third of all tool results ($3{,}827$ of $11{,}189$) were cut before they
reached the model, and every cut result became broken JSON. This loss
followed a clear pattern. The harness cut each result by slicing its JSON
text at a fixed length. A slice like this does not shorten a result evenly.
It deletes whatever data sits near the end. One field reached the model in
only $4.7\%$ of calls. A second field, in the very same results, reached the
model in $99.3\%$ of calls. Which field held the fault's evidence depended
on when the faulty process started: early or late in the trace. The agent's
diagnosis accuracy matched this same split, almost exactly. So we ran
controlled experiments to test one question: did the data loss
\emph{cause} the drop in accuracy? This paper makes four contributions.
First, we model an agent's context as a five-stage pipeline, with a
measured failure at each stage. Second, we replay $864$ real tool results,
including one from a production trace where the shipped cut deletes all
$128$ records of the fault's evidence. Third, a controlled experiment shows
that the cut's real cost appears first as extra work, not as wrong answers:
$43\%$ more tool calls, twice the computation, and yield falling from
$0.61$ to $0.31$. The agent still finds the right answer when it has
another way to reach it. Fourth, a survey of seven agent frameworks finds
two that cut tool output the same unsafe way, with no consistent, safer
alternative in the rest. Whether the data cut truly causes the accuracy
drop is still an open question. A repeated test on real production data
shows the same pattern again, but it does not settle the question.
\end{abstract}

\begin{IEEEkeywords}
LLM agents, agent harnesses, context engineering, empirical evaluation
\end{IEEEkeywords}

\section{Introduction}
\label{sec:intro}

An agent never reads its context window directly. It reads whatever its
harness, the code that wraps the model, decided to put there. Between a
tool producing a result and the model seeing it, the harness serializes
it, cuts it to a budget, keeps or drops parts across turns, and passes it
between phases. Every one of these steps is a design choice. Every one of
them can lose information.

Most research on agent context asks how big the window is and how the
model attends within it~\cite{liu2024lost,hsieh2024ruler}, or how to extend it with
memory~\cite{packer2023memgpt}. Work on agent harnesses often treats the surrounding code as a minor
detail~\cite{yang2024sweagent,jimenez2024swebench}, a gap also noted
elsewhere~\cite{lindenbauer2025complexity}.
Two recent studies read the source code of eleven~\cite{harness2026} and
thirteen~\cite{rombaut2026scaffold} production harnesses, but neither
measures what happens to the data. Failure
taxonomies label what went wrong across thousands of agent
runs~\cite{cemri2025mast}, but a label is not a loss measurement. Four recent studies come closer to ours. One asks how well a model uses a
tool's \emph{complete} output~\cite{kate2026tooloutputs}; we ask whether
the output is complete in the first place. One measures how a harness's
design changes outcomes across models and
tasks~\cite{yao2026harnessbench}; we trace that variance to one measurable loss point, not the outcome.
One varies context-management policies and window budgets in a coding
harness~\cite{fan2026harnessdesign}; it reports success and cost, but not
which evidence each policy deletes. One shows that storing tool
output in a file for retrieval, instead of context, costs accuracy on
some models but not others~\cite{zhang2026samesimulation}; our cause is
different: destructive delivery, not a retrieval choice. The
gap all of this work shares: a run log records what a tool
\emph{delivered}, not what it \emph{produced}. A loss on the way into the
log cannot be found in the log.

We measured this gap in an agent that finds the root cause of injected
faults from Linux kernel traces. Such agents are an active research
area~\cite{chen2024rcacopilot,xu2025openrca}. Three facts came out of that
measurement.

First, the window was nearly empty. Across all $720$ runs, the single
largest prompt we sent was only $23$k tokens, against a $400$k window.
That is $5.75\%$.

Second, a third of all tool results were destroyed before they reached
the model. The harness did this by cutting long results at a fixed point
in the text. JSON always lists its keys in the same order, so a cut like
this does not shorten a result evenly, it deletes whichever keys come
later, completely. One list of containers reached the model in only
$4.7\%$ of $742$ calls. A second list, sitting right next to it in the
same payload, reached the model in $99.3\%$ of calls.

Third, the damage lined up exactly with the task we asked the agent to
solve. A fault that \emph{starts} a new process during the trace shows up
in the early list, the one that survives. A fault in a container that was
\emph{already running} shows up only in the late list, the one that dies.
Diagnosis accuracy split along that same line, on both test applications:
faults that spawn a new process scored $40$--$55$ out of $60$; faults in
containers already running scored $0$--$24$ out of $60$. We used to think
this gap was just a limit of what kernel traces can show. The delivery
numbers say otherwise: our own harness is partly responsible. How large a
part is still an open question. We report what our experiments settled,
and what they did not.

Our contributions are four. (1) A five-stage pipeline model of an agent's
context, with a measured failure at each stage, and an instrument to
measure it: per-stage counts and one \emph{yield} number, applied to our
own harness (Sec.~\ref{sec:pipeline}). (2) Replay evidence, including a
case where the shipped cut deletes an entire evidence class at
production scale (Sec.~\ref{sec:measured}). (3) A registered, controlled
experiment whose result is a \emph{compensation tax}: when the agent has
another way to reach the answer, the cut costs it work, not accuracy
(Sec.~\ref{sec:ablation}). (4) A survey of seven agent frameworks: two of
them cut tool output the same unsafe way, and the rest offer no
consistent, safer alternative (Sec.~\ref{sec:measured}).

\section{Context as a Pipeline}
\label{sec:pipeline}

Picture one diagnostic run. In the usual reason--act
loop~\cite{yao2023react}, the agent calls a tool, say a container
listing, and gets a result back (\emph{produced}). The harness turns
that result into text and fits it inside a budget before the model ever
reads it (\emph{delivered}). The agent splits its investigation across
several \emph{workers}, sub-tasks with their own limit on tool-calling
turns; a worker acts on what it receives, turn by turn (\emph{act}), and
keeps only so much conversation history before older turns get dropped
(\emph{retain}), a separate limit of its own. A worker that computes
something worth reporting must
hand it off to whatever writes the final verdict (\emph{transfer}); if
it does not, the computation is lost even though it happened. Nothing
here carries to the next run (\emph{persist}); we turn that off by
design, to keep runs independent. We model this five-step path as five
boundaries (Table~\ref{tab:pipeline}), which fail in different ways that
matter for how to read it. S1 and S4 destroy evidence outright: a result or
a finding is simply gone, as if it never existed. S2 does not destroy
anything; it cuts a worker off before it finishes, so some work never
happens at all. S3 does not destroy anything either; it only makes a run
more expensive, by resending history that did not need resending, an open
design question~\cite{lindenbauer2025complexity,fan2026harnessdesign}. S5
looks absent in our setup, but that is a choice, not a failure: we turned
it off on purpose, so one run cannot leak into the next. We still call
all five one pipeline, because each is a point where the harness decides
what survives to the next step. That is what they share. How they fail is
not it.

The loss at a boundary is not random: it depends on how the cutting rule
interacts with the shape of the data. If a result keeps a fixed key
order and the cut is a prefix slice, whether a field survives depends
only on its \emph{position} in the text. Nothing makes a late field less
important than an early one, yet models already attend unevenly to
position inside the window~\cite{liu2024lost}. The harness makes this
worse: a field past the cut is not attended to less, it is gone before
attention ever gets the chance to see it.

\begin{table}[t]
\caption{The five boundaries, and the failure found at each one. S1--S4
come from the harness's own field data, collected \emph{before} the
controlled experiment in Sec.~\ref{sec:ablation}; S5 is a design choice,
not a measured failure. These rows show how often each boundary fails.
Whether that failure changes the final answer is a separate question,
tested only by the controlled experiment.}
\label{tab:pipeline}
\centering\small
\begin{tabular}{@{}llp{4.0cm}@{}}
\toprule
& \textbf{Boundary} & \textbf{Measured failure (field data)} \\
\midrule
S1 & produce $\to$ deliver & $720$-run eval: prefix slicing, $34\%$ of results cut, all broken; one field delivered $4.7\%$ vs $99.3\%$ \\
S2 & deliver $\to$ act & $554$-worker campaign: the step budget is a wall, $51\%$ of workers stopped at exactly the limit \\
S3 & act $\to$ retain & one before/after run: raising retention $80$k to $150$k chars moved it from $303$k to $545$k tokens \\
S4 & retain $\to$ transfer & same $554$-worker campaign: hand-off is voluntary, $98$ workers passed on nothing; silence reads as ``nothing found'' \\
S5 & transfer $\to$ persist & by design, not measured: nothing survives the run (to keep runs independent) \\
\bottomrule
\end{tabular}
\end{table}

Table~\ref{tab:pipeline}'s numbers answer only \emph{how often} each
boundary fails, not \emph{whether} a failure changes the final answer; a
correlation in field data does not prove that. Section~\ref{sec:ablation}
answers that question directly, by re-introducing one boundary's failure
at a time, with everything else held fixed.

\textbf{Measuring the loss.} We propose instrumenting these boundaries
and reporting, next to accuracy, how much produced evidence survived
each one, with a concrete unit at each boundary. Others argue for reporting
cost, not just accuracy~\cite{kapoor2025agents}, and where a chain
breaks~\cite{zhang2026phreeqc}; we add a loss count at each boundary. At S1, the unit is a
top-level field of a tool result, counted as delivered when its content
is present in the text the model receives (whether the payload still
parses is reported separately). At S4, the unit is a recorded finding or
a captured computation, counted as transferred when it reaches the phase
that decides. Our summary number, \emph{yield}, is findings recorded
divided by computations run: a narrow, measurable proxy for ``work that
reached the decision,'' not for relevance, which needs ground truth and
is future work.

\section{The Defect, Measured}
\label{sec:measured}

\subsection{In the field}
We now give the full numbers behind Table~\ref{tab:pipeline}'s S1 row,
from the harness's deployed evaluation: $720$ runs, $11{,}189$ tool
results. Of those, $3{,}827$ were cut, and every one broke. Inside the
cut results, one field reached the model in only $4.7\%$ of $742$ calls;
a second field, in the same results, reached it in $99.3\%$.

This is a separate point from accuracy: the scores we give elsewhere
(the $40$--$55$ versus $0$--$24$ split in the introduction) come from a
strict scorer, which only counts an answer correct when the agent names
the right container against ground truth, not from an earlier, lenient
scorer that counted any mention in passing. We caught that scorer's
mistake and replaced it before running our controlled experiment.

\subsection{Mechanism and replays}
\label{sec:e1}

\begin{figure}[t]
\centering
\includegraphics[width=\columnwidth]{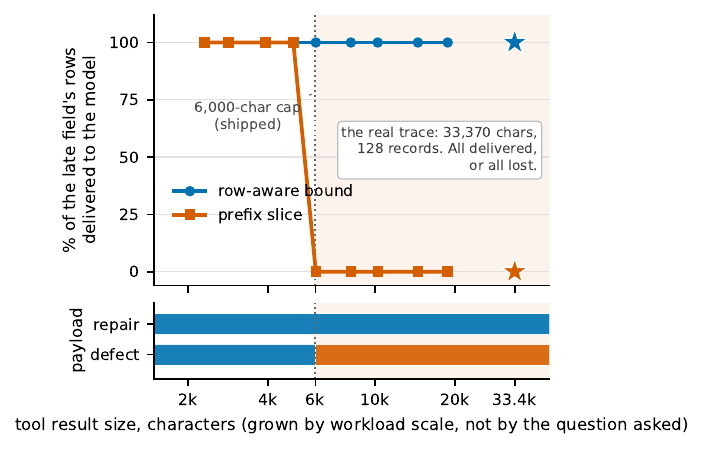}
\caption{The delivery cliff. We build a result with the real tool's shape
and replay it through both cutting rules as it grows past the shipped
$6{,}000$-character cap. No model is involved. Under the prefix slice, the
late field arrives whole until the cap, then vanishes entirely and the
payload breaks (bottom strip). The row-aware cut keeps every field at every
size. The stars mark where the real production trace (Sec.~\ref{sec:e1})
falls: $33{,}370$ characters, $128$ records, delivered or lost whole,
exactly as predicted.}
\label{fig:cliff}
\end{figure}

Section~\ref{sec:pipeline} explained \emph{why} a prefix slice loses data
by position. Here we show that this is not just a theory: it is
deterministic, and it holds on real, recorded data (Fig.~\ref{fig:cliff}).
For every tool call, our transcripts store both the full result and the
text actually sent to the model. We replayed $864$ recorded results,
from $31$ runs, through both cutting rules: the shipped prefix slice and
the repaired row-aware cut. Every result over the budget broke under the
prefix slice, every time; none broke under the row-aware cut. The tools
that overflow the budget are the ones that \emph{list things}: $50\%$ of
process-listing calls broke, along with $43\%$ of the agent's own
analysis outputs and $1\%$ of small summaries. Only $2.5\%$ of all calls
were affected at all, which is exactly why this defect survives
inspection: quiet on most calls, fatal on the few that matter.

We then replayed the shipped cut on one production-scale trace: a real
$21$-container deployment, with one service held under a CPU limit. The
culprit here is what we later call \emph{quiet} (Sec.~\ref{sec:ablation}):
already running before the trace starts, with no second, independent way
for the agent to spot it. It is one Java container among five, each
close to $1.4\%$ of kernel events, so it does not stand out by size
either. The listing result for this trace is $33{,}370$ characters,
$5.6\times$ the shipped budget; we tuned nothing, that is simply how big
a real result gets. The shipped slice delivers $30$ of $118$ early-list
records, but \textbf{0 of 128} late-list records, in an already
broken payload. On real deployed data, at the real budget, this
culprit's entire class of evidence disappears before the model sees it.

\subsection{The same cut in other frameworks}
A defect found in a single harness would be a narrow result. To test how
general it is, we examined the tool-output cutting code in current
releases of seven agent frameworks (langchain-core 1.6.12, langchain
1.4.6, smolagents 1.26.0,
openai-agents 0.24.0, autogen-agentchat 0.7.5, pyautogen 0.10.0, crewai
0.11.2). We found two general-purpose cutters, both slicing by position,
just like ours. langchain-core slices serialized tool output down to
$500$ characters, keeping only the front; smolagents deletes the
\emph{middle} of any observation over $20$k characters. Neither looks at
the structure of what it cuts. Three frameworks do not cut tool output
at all, which just moves the same loss further along, into the context
window instead. One framework cuts only its own tracing data. We found
one structure-aware cutter: a shell middleware that cuts by whole lines
and says so. It shows the safe pattern exists. It is just not the
default.

\section{The Controlled Experiment}
\label{sec:ablation}

The field numbers show that delivery and accuracy move together: a
match, not a cause. There is also a confound: when we fixed the delivery
boundary, we also shipped two new analysis tools, so an accuracy gain in
the field data could come from those tools, not the fix. To separate
cause from confound, we put each defect back on its own, one at a time,
holding the agent, tools, task, and model fixed. We wrote down our
predictions before each wave of runs, echoing registered
reports~\cite{ernst2023registered}. Every change to the plan is
logged, with its date and reason, in a file that ships with the
artifact.

\textbf{Setup.} The subjects are kernel traces of a coding agent running
under resource limits, with exact ground truth. The diagnosing agent,
model (temperature $0$), tools, and task all stay fixed. There are two
fault classes: in \emph{spawned} faults, the culprit process starts
partway through the trace; in \emph{resident} faults, it was already
running. Healthy traces serve as negative controls, and every run gets
the same one-line symptom report.

We test six configurations. \emph{Repaired} is the fixed version:
row-aware cut, automatic hand-off. \emph{S1 defect} re-introduces just
the prefix-slice cut, at the budget that bites on these traces.
\emph{S1 reversed} keeps that defect, but serializes keys in reverse
order, so only position changes, not content. An \emph{order-only}
control reverses keys with no cut, isolating the effect of reordering
alone. Single-defect arms re-introduce S2 through S4 one at a time.
\emph{All defects} re-introduces all five together, matching the
original shipped harness.

Each fault class has $5$ distinct recordings, times $3$ repeats (temperature
$0$ is not fully repeatable~\cite{ouyang2025nondeterminism}), paired across configurations: $145$ runs total, none failed. Scoring counts
only a \emph{commitment}: the agent must name the culprit in its answer
field, checked against ground truth, not a mention in passing. Our
first scorer counted mentions instead, inflating one arm's score from
$2/15$ to $14/15$; we caught the mistake auditing at $46$ runs and
replaced the scorer.

\begin{table}[t]
\caption{The cut changes the price, not the answer. Medians over the $15$
spawned-culprit runs per configuration ($5$ recordings $\times$ $3$ repeats;
order-only has $10$). Yield = findings recorded $\div$ computations run.}
\label{tab:compensation}
\centering\scriptsize
\setlength{\tabcolsep}{3pt}
\begin{tabular}{@{}lrrrrrr@{}}
\toprule
 & \textbf{correct} & \textbf{calls} & \textbf{cut/run} & \textbf{comp.} & \textbf{yield} & \textbf{tokens} \\
\midrule
repaired         & 14/15 & 53 & 0  & 18 & 0.61 & (567k) \\
S1 defect        & 15/15 & 68 & 27 & 32 & 0.34 & $+5\%$ \\
S1 reversed      & 15/15 & 76 & 30 & 36 & 0.31 & $+14\%$ \\
order only       & 10/10 & 48 & 0  & 16 & 0.61 & $-12\%$ \\
\bottomrule
\end{tabular}
\end{table}

\textbf{Results} (Table~\ref{tab:compensation}). The window was never the
limit: a run used a median $567$k input tokens, over a median $53$
calls, about $10$k per call, just $2.5\%$ of the $400$k window at that
median. Under the defect, the agent lost a median $27$ results per run,
$40\%$ of its calls, every one broken. It still
named the spawned culprit in $15/15$ runs, against $14/15$ for the
repaired condition. That difference is noise at this sample size, not a
sign the defect helps. What moved was the work. From repaired to S1
defect to S1 reversed, tool calls rose from $53$ to $68$ to $76$: a
$28\%$ rise, then a $43\%$ rise. Computations rose the same way, from
$18$ to $32$ to $36$, roughly doubling. Input tokens rose $5\%$, then
$14\%$. Yield fell from $0.61$ to $0.34$ to $0.31$, because the agent
runs more computations without finding more, not because it finds less:
findings stay at $11$ throughout. The order-only control, changing only
key order and cutting nothing, lands back at $0.61$ too, confirming
yield measures the real effect, not reordering. At
equal answers, the extra tokens are unnecessary inference cost~\cite{kapoor2025agents} that
will not show up in end-to-end latency.

\textbf{The reversal test, and a refuted prediction.} We predicted that
reversing the key order would move the culprit's list from fully
delivered to fully deleted, so accuracy should drop. The manipulation
worked as planned: transcripts show the culprit absent from every
listing payload. Yet the agent still named it, $15/15$. It recovered the
name differently: the culprit, the loudest process in the trace at
$14.0\%$ of kernel events, lands in the first rows of every sorted table
the agent builds, the part that survives. The order-only control stayed flat,
so this recovery is real, not an artifact of the scoring. Our original
prediction was wrong. What replaces it is a sharper rule: \textbf{the cut
hurts only where the agent has no second route to the answer, and its
cost shows up as extra work before it shows up as errors.} A loud culprit
survives the loss of any single route. A quiet culprit that lives in one
field, like the resident case in the field data, delivered at just
$4.7\%$, has no second route.

\textbf{A direct replay, repeated, at production scale.} Three early
attempts to show the cut destroying \emph{accuracy}, not data, failed:
with no working repaired baseline there was nothing to compare against.
We then fixed two things: delivery, so results are never truncated
mid-record, and the step budget, giving workers more tool-calling turns
before stopping. We also built a page-on-request mechanism, a safety net
for oversized results that never fired here, since every result fit on
the first page. Other work also finds such recall tools rarely
used~\cite{fan2026harnessdesign}. We reran the same trace and prompt under both
conditions. Repaired: $3$
of $6$ runs name the throttled service correctly, $0$ of $6$ broken
payloads. S1 defect: $0$ of $5$ correct, every run broken, $10$ to $14$
payloads each. The data-destruction result holds with no exceptions at
production scale too. The accuracy half moves the same direction but is
noisier here: three of six repaired runs were still wrong, for reasons
unconnected to data loss. We read this as a direct, repeated replication
of the mechanism at deployed scale, not proof the cut alone drives the
field-wide split.

\section{Why Nobody Notices}
\label{sec:unnoticed}

Three things hide this class of defect. First, it costs almost no time:
in kernel traces from two production coding agents ($10$ uncapped runs),
waiting on the model ate a median $84\%$ of wall clock (range
$80$--$91\%$, network and queueing included). The defect's real cost, in
correctness and token spend, barely touches that metric, the one teams
actually watch. Second, capacity looks fine: a window barely touched
gives no hint that evidence is being discarded upstream. Third, the
usual remedy hides the symptom: handing the agent a written procedure
reached the same verdict as free exploration, at just $35\%$ of the
tokens. Teams that feel context pressure add
structure, see an improvement, and stop looking there. It is the right
fix for the wrong reason: it needs less delivered evidence, instead of
delivering evidence better.

\section{Threats to Validity}

One harness, one domain, one model: we argue the defect class is
general because the pipeline itself is unavoidable, and two of the
seven surveyed frameworks cut the same way, with no consistent
alternative (Sec.~\ref{sec:measured}). The same shape recurs whenever a harness
truncates a compiler error, a CI log, or a failing test's output before a coding agent gets to see it; one coding harness caps every tool
result at $24$k characters~\cite{fan2026harnessdesign}. We have not measured other harnesses end to
end, nor a conventional build-test-fix coding agent; the evidence here
is kernel traces only. We also wrote the defect ourselves, so the
experiment reintroduces it under registered predictions, not as a repair
story. On scale: $5$ recordings per class, with paired repeats of the
same five recordings, not five independent fault instances. The
experiment's budget is scaled to fit its smaller traces; the
production-scale replay keeps the shipped budget. The accuracy half of
the field split remains correlation, not cause.

\section{Future Plans}

\textbf{(1) The causal accuracy experiment.} Pick a fault family with a
known-succeeding repaired agent and a quiet culprit. Hold the trace,
fault, prompt, model, and tools fixed, varying only delivery: complete,
prefix-truncated, reversed-key-order, row-aware, and paging-on-request
(not yet seen to fire). For every run, report whether the culprit's
records are visible, whether the payload is valid, whether the culprit
is recoverable elsewhere, and the final answer, across independent
instances.
\textbf{(2) A deeper survey.} Extend our reading to the eleven-harness
corpus of~\cite{harness2026} and the thirteen scaffolds
of~\cite{rombaut2026scaffold}, classify each cutter as position- or
structure-based, and replay it against the delivery cliff.
\textbf{(3) Reporting loss as standard practice.}
Package the instrument so any harness can print its loss profile beside
its accuracy number, to test whether equal-accuracy harnesses differ in
yield and predict harder-task performance.
\textbf{(4) Tests without a model.} Each failure in Table~\ref{tab:pipeline}
is mechanical, so CI can catch it: a cut result that fails, a step
budget at its limit, a dropped computed result. A replication package
(drivers, scorer, results, replay corpus, registered plan) is included
(see footnote on page 1).

\section{Conclusion}

An agent's context is not a window: it is a pipeline's output. That
pipeline destroyed a third of all tool results while the window sat
$93\%$ empty, deleting evidence by position, not value. Where the
agent had another route, the loss cost extra work, not accuracy. Every
harness makes these choices and none reports them. They are easy to
measure. We have shown how.

\balance

\end{document}